\documentclass[usenatbib,useAMS]{mn2e}
\usepackage{epsfig}
\usepackage{times}
\usepackage{epsfig}
\usepackage{graphicx}
\usepackage{bm}
\usepackage{dcolumn}
\usepackage{graphicx}
\usepackage{epstopdf}
\usepackage{color}
\usepackage{epsf}
\usepackage{epsfig}
\usepackage{graphicx,epic,eepic,color}

\begin{document}

\title[Testing MOG with Chandra X-ray Clusters]{The MOG Weak Field approximation II. Observational test of Chandra X-ray Clusters}
\author[J. W. Moffat and S. Rahvar]
{J. W. Moffat$^{1,2}$\thanks{jmoffat@perimeterinstitute.ca}, S. Rahvar$^{1,3}$\thanks{rahvar@sharif.edu}  \\
$^1$ Perimeter Institute for Theoretical Physics, 31 Caroline St. N., Waterloo, ON, N2L 2Y5,Canada  \\
$^2$ Department of Physics and Astronomy, University of Waterloo,
Waterloo, Ontario N2L 3G1, Canada\\
$^3$ Department of Physics, Sharif University of Technology, P.O.
Box 11155-9161, Tehran, Iran }

\maketitle

\begin{abstract}
We apply the weak field approximation limit of the covariant Scalar-Tensor-Vector Gravity (STVG) theory, so-called MOdified gravity (MOG), to the dynamics of clusters of galaxies by using only baryonic matter. The MOG effective gravitational potential in the weak field approximation is composed of an attractive Newtonian term and a repulsive Yukawa term with two parameters $\alpha$ and $\mu$. The numerical values of these parameters have been obtained by fitting the predicted rotation curves of galaxies to observational data, yielding the best fit result: $\alpha = 8.89 \pm 0.34$ and $\mu= 0.042\pm 0.004$ kpc$^{-1}$~\cite{rah13}. We extend the observational test of this theory to clusters of galaxies, using data for the ionized gas and the temperature profile of nearby clusters obtained by the Chandra X-ray telescope. Using the MOG virial theorem for clusters, we compare the mass profiles of clusters from  observation and theory for eleven clusters. The theoretical mass profiles for the inner parts of clusters exceed the 
observational data. However, the observational data for the inner parts of clusters (i.e., $r<0.1 r_{500}$) is scattered, but at distances larger than $\sim 300$ kpc, the observed and predicted mass profiles converge. Our results indicate that MOG as a theory of modified gravity is compatible with the observational data from the the solar system to Mega parsec scales without invoking dark matter.\\


\end{abstract}


\section{introduction}

Zwicky showed that there is an inconsistency between the dynamical mass and the luminous mass for the Coma cluster. He postulated that there must  
exist dark matter at the scale of clusters of galaxies~\cite{wicky}.  The observations of the rotation curves of spiral
galaxies by Rubin also indicated that galaxies need missing mass in the context of Newtonian gravity~\cite{rubin}. The main paradigm to interpret missing mass in the universe is the theory of dark
matter, supposing that the unseen matter accounts for more than $80 \%$ of the 
universe's matter content. This hypothetical matter is made of
particles (e.g.,WIMPs) that may only interact weakly with ordinary baryonic matter.
Ongoing experiments have been conducted to directly
detect dark matter particles by sensitive particle physics detectors. So far, no
positive signal has been detected~\cite{wimp,lux}.

An alternative approach for solving the problem of missing mass is to modify the law of gravity. One of the first phenomenological 
efforts to modify Newtonian gravity was MOdified Newtonian Dynamics (MOND)~\cite{mond}. This model is based on a non-linear Poisson equation for the gravitational potential with a characteristic acceleration $a_0$; Newtonian gravity is modified for particles with acceleration $a < a_0$. The model can successfully describe the rotation curves of galaxies~\cite{mond}, but it fails to be consistent with the temperature and gas density profiles of clusters of galaxies without invoking dark matter~\cite{angus}.  There are other modified gravity models such as the relativistic extension of MOND by Bekenstein (2004), conformal gravity~\cite{conformal} and non-local gravity~\cite{bahram,hehl}.

We attempt in the following to test observationally a covariant MOdified Gravity (MOG) theory, also 
called Scalar--Tensor--Vector Gravity (STVG) at the scale of clusters of galaxies~\cite{moffat06,mt1,mt2,mt3,b1,b2,b3}.  In this theory, the dynamics of a test particle are given by a modified equation of motion where, in addition to the conventional geodesic motion of a particle, a massive vector field couples to the fifth force charge of a particle. The fifth force charge of a particle is proportional to the inertial mass of the particle. We have shown that in the weak field approximation the potential for a matter distribution of an extended object behaves like the Newtonian potential with an enhanced gravitational constant and an additional Yukawa potential~\cite{rah13}. We applied the effective potential in the weak field approximation to the study of the dynamics of galaxies and demonstrated that MOG can successfully explain the rotation curves of galaxies without the need for dark matter. We fixed the two free parameters of the potential and chose the stellar mass-to-light ratio as the only free parameter in calculating the dynamics of the galaxies. 

We extend in the following the observational tests in Moffat \& Rahvar (2013) to one order of magnitude larger system scale. We 
use the observational data of relaxed clusters of galaxies in the Chandra X-ray sample for this purpose. For a set of clusters of galaxies, we have the temperature profile as well as the plasma gas and galaxy distribution. These observables are sufficient to calculate the dynamical 
mass and compare it to the baryonic mass of clusters.  In section (\ref{formalism}), we briefly introduce the 
STVG theory and review the weak field approximation limit. In section (\ref{sectionBoltzmann}), we obtain a full relativistic version of the Boltzmann equation in MOG. Furthermore, we derive the virial equation in the weak field approximation. In section (\ref{data}), the observational data from the Chandra X-ray sample of clusters are examined in conjunction with the MOG and finally in (\ref{conc}), we provide a summary and conclusions.

\section{Scalar-Tensor-Vector Gravity (STVG): Weak Field Approximation}
\label{formalism}
The STVG theory is a covariant  theory of gravity. The action of STVG is composed of scalar, tensor and vector fields~\cite{moffat06}:
\begin{equation}
\label{action1}
S=S_G+S_\phi+S_S+S_M.
\end{equation}
The components of the action are (i) the Einstein gravity action:
\begin{equation}
S_G=-\frac{1}{16\pi}\int\frac{1}{G}\left(R+2\Lambda\right)\sqrt{-g}~d^4x,
\end{equation}
where $\Lambda$ is the cosmological constant, (ii) the massive vector field $\phi_\mu$ action:
\begin{eqnarray}
S_\phi&=&-\frac{1}{4\pi}\omega\int\Big[\frac{1}{4}B^{\mu\nu}B_{\mu\nu}-\frac{1}{2}\mu^2\phi_\mu\phi^\mu\nonumber\\
&+&V_\phi(\phi_\mu\phi^\mu)\Big]\sqrt{-g}~d^4x,
\end{eqnarray}
where $\phi^\mu$ is the vector field and $B_{\mu\nu} = \partial_\mu\phi_\nu -  \partial_\nu\phi_\mu$, (iii) the action for the scalar $G$ and $\mu$ fields is:
\begin{eqnarray}
S_S&=&-\int\frac{1}{G}\Big[\frac{1}{2}g^{\alpha\beta}\biggl(\frac{\nabla_\alpha G\nabla_\beta G}{G^2}
+\frac{\nabla_\alpha\mu\nabla_\beta\mu}{\mu^2}\biggr)\nonumber\\
&+& \frac{V_G(G)}{G^2}+\frac{V_\mu(\mu)}{\mu^2}\biggr]\sqrt{-g}~d^4x,
\label{scalar}
\end{eqnarray}
where $\nabla_\mu$ denotes the covariant derivative with respect to the metric $g_{\mu\nu}$. $\omega$ denotes a dimensionless coupling constant and 
$V_\phi(\phi_\mu\phi^\mu), V(G)$ and $V(\mu)$ denote self-interaction potentials of $\mu$ and $G$ 
fields.  While in the generic formalism of MOG, $G$ and $\mu$ are taken into account as scalar fields, for simplicity we keep these two fields as constant parameters. Finally, (iv) $S_M$ is the matter action.

The action for pressureless dust coupled to the vector field $\phi_\mu$ is given by
\begin{equation}
S_M = \int(- \rho \sqrt{u^\mu u_\mu} - \omega Q_5 u^\mu\phi_\mu)\sqrt{-g} dx^4,
\label{SM}
\end{equation}
where the density of the fifth force charge $Q_5$ is related to the 
density of matter by $Q_5 = \kappa \rho$ where $\kappa$ is a constant. 
For a test particle with the density $\rho(x) = m\delta^3(x)$ and a fifth force charge $q_5=\kappa m$, a variation of the action in Eq. (\ref{SM}) with respect to the comoving time $\tau$, yields the equation of motion:
\begin{equation}
\frac{du^\mu}{d\tau} + \Gamma^\mu{}_{\alpha\beta}u^\alpha u^\beta =
\omega\kappa B^\mu{}_\alpha u^\alpha, 
\label{geo}
\end{equation}
where the right-hand side is a Lorentz-type force. Since the fifth force charge is proportional to the inertial mass, the equation of motion is independent of the mass. The main difference of the Lorentz-type force with standard electromagnetism is the same 
sign for the charge of particles, unlike the normal electromagnetic force where there are opposite sign particles (results in a neutral medium for an ensemble of particles in a plasma). A physical consequence of the same sign charged particles is a repulsive force produced by an ensemble of particles. Since the vector field in MOG is massive, after a characteristic range of the field given by $\l = 1/\mu$, the effect of the repulsive force vanishes and Eq. (\ref{geo}) reduces to the pure metric field geodesic equation of motion.

\subsection{Weak Field Approximation}
For the astrophysical application of MOG, we need the weak field approximation of the theory. We briefly review in this section the formalism of the weak field approximation.  A detailed derivation of this limit for extended distributions of matter was presented in Moffat \& Rahvar (2013). 

For a slow moving particle, we expand the metric around Minkowski space:
\begin{equation}
g_{\mu\nu} = \eta_{\mu\nu} + h_{\mu\nu}.
\end{equation}
The left-hand side of Eq. (\ref{geo}) can be written in terms of $h_{00}$, while on the right-hand side we can ignore the magnetic-like force compared to the electric-like force due to the slow motion of the particle.  Eq. (\ref{geo}) can now be written as
\begin{equation}
\ddot{x}^i =-\partial^i\biggl(-\frac{1}{2}h_{00}-\omega\kappa\phi^0\biggr),
\label{geo2}
\end{equation}
where $i=1,2,3$ and we define the effective potential as 
\begin{equation}
\Phi_{eff} =-\frac{1}{2} h_{00}-\omega\kappa\phi^0. 
\end{equation}
We substitute in the right-hand side of (\ref{geo2}) the solutions for $h_{00}$ and $\phi^0$ from the field equations. For the tensor component, $\Phi_N = -\frac12 h_{00}$ where $\nabla^2\Phi_N = 4\pi G\rho$, and the zero component of the vector field $\phi^0$ satisfies the Yukawa equation:
\begin{equation}
\nabla^2\phi^0 - \mu^2\phi^0 = -\frac{4\pi}{\omega}J^0,
\label{phi}
\end{equation}
where $J^0 = Q_5$ is the zero component of four dimensional current as the source of vector field and relates to density of 
matter as $Q_5 = \kappa \rho$. Substituting the solution of the Poisson equation and the solution of $\phi^0$ from the Yukawa equation into (\ref{geo2}) and taking into account the 
asymptotic solutions of effective potential at smaller distances (i.e. $r\ll \mu^{-1}$) to the Newtonian 
gravity \cite{rah13}, we obtain the following effective potential in the weak field approximation:
\begin{eqnarray}
\Phi_{eff}(\vec{x}) &=& - G_N(1+\alpha) \int\frac{\rho(\vec{x}')}{|\vec{x}-\vec{x}'|} d^3x' \nonumber \\
 &+&G_N \alpha \int\frac{\rho(\vec{x}')}{|\vec{x}-\vec{x}'|} 
e^{-\mu|\vec{x}-\vec{x}'|}d^3x', \label{potential22}
\end{eqnarray}
where $G$ has been replaced by $G = G_N (1+\alpha)$, $\alpha$ is a dimensionless 
parameter, $G_N$ is the Newtonian gravitational constant and $\mu$ is the mass of the vector field. We have used this effective potential in Moffat \& Rahvar (2013) and successfully fitted the rotation curves of galaxies, yielding the best values of the parameters:
$\alpha =8.89 \pm 0.34$ and $\mu =0.042 \pm 0.004$ kpc$^{-1}$.

The first term on the right-hand side of Eq. (\ref{potential22})
is the potential of the Newtonian attractive force with the effective gravitational 
constant $G=G_N(1+\alpha)$ and the second term results from the repulsive vector force. For scales much larger than $\mu^{-1}$, the Yukawa term is suppressed and the gravitational potential is enhanced by factor of $1+\alpha$ compared to the Newtonian potential. On the other hand, for smaller scales the repulsive force cancels the enhanced attractive force and the result is the standard Newtonian potential at solar system scales.

\section{Boltzmann Equation in MOG}
\label{sectionBoltzmann}

In this section we derive the full relativistic version of the Boltzmann equation in MOG and for the astrophysical applications, we calculate the Boltzmann equation in the weak field approximation.  The result will be used in studying the dynamics of X-ray clusters of galaxies. 

For an ensemble of collisionless particles moving due to their self-gravity, the number of particles inside a
differential element in the phase-space does not change and ${D f}/{d\tau}=0$. This equation can be written as a sum of
partial derivatives in terms of time, space and momentum. Since in
momentum space, we have the constraint $p^\mu p_{\mu} = m^2$,
there are three degrees of freedom for $p^\mu$ and we keep only the
spatial components of the momentum space in our calculation~\cite{lin64}. The Boltzmann equation can be written as
\begin{equation}
{\dot x^\mu}\frac{\partial f}{\partial x^\mu} +
\frac{dp^i}{d\tau}\frac{ \partial f}{\partial p^i} = 0. \label{1}
\end{equation}
We can write (\ref{geo}) in terms of the momentum of a particle:
\begin{equation}
\frac{dp^i}{d\tau} = -\frac{1}{m}\Gamma^i_{\mu\nu}p^\mu p^\nu +
\kappa\omega  B^i{}_{\mu} p^\mu . \label{3}
\end{equation}
By substituting (\ref{3}) into (\ref{1}), we obtain the Boltzmann equation in curved space with the extra vector field as follows:
\begin{equation}
p^\mu\left[\frac{\partial}{\partial x^\mu}-(
\Gamma^i{}_{\mu\nu}p^\nu-{\kappa}\omega m
B^i{}_\mu)\frac{\partial}{\partial p^i}\right]f = 0.
\end{equation}

In order to have a relativistic Euler equation, we multiply the
Boltzmann equation by the spatial component of four-momentum, $p_j$,
and integrate over momentum space:
\begin{equation}
\int p^\mu p_j \frac{\partial f }{\partial x^\mu} dV_p -\int p^\mu
p_j  (p^\nu \Gamma^i{}_{\mu\nu}- \kappa \omega m
B^i{}_\mu)\frac{\partial f}{\partial p^i} dV_p = 0, \label{jeans1}
\end{equation}
where $dV_p$ is the differential volume in momentum space. In the first term, 
the space-time partial derivative can be moved outside the integral, and in the second term 
the Christoffel symbol and the vector field strength can also be moved outside the integral. Integrating
over momentum space gives the result:
\begin{eqnarray}
\frac{\partial}{\partial t}(\rho \overline{p^0p_j}) &+&
\frac{\partial}{\partial x^i}(\rho\overline{p^ip_j}) +
\delta_{ij}\Gamma^i{}_{\mu\nu}\rho\overline{p^\mu p^\nu}\nonumber \\
 &+& 2 \Gamma^i{}_{i\mu}\rho \overline{p^\mu p_j}- \delta_{ij} \kappa \omega m\rho B^i{}_{\mu}\overline{p^\mu}= 0,
\label{jeans}
\end{eqnarray}
where the bar sign denotes the average value over momentum space. For
simplicity we rewrite the relativistic Euler equation in terms of
the four velocity $u^\mu$:
\begin{eqnarray}
\frac{\partial}{\partial t}(\rho g_{j\beta} \overline{u^0 u^\beta})
&+& \frac{\partial}{\partial x^i}(\rho\overline{u^iu_j}) +
\delta_{ij}\Gamma^i{}_{\mu\nu}\rho\overline{u^\mu u^\nu} \nonumber \\
&+& 2 \Gamma^i_{i\mu}\rho \overline{u^\mu u_j} - {\kappa} \omega
\rho \delta_{ij}B^i{}_{\mu}\overline{u^\mu}= 0. \label{b11}
\end{eqnarray}
Now we multiply this equation by $x^k$ and integrate over the 3-space
volume and calculate this expression term by term. Using the
definition of the inertia tensor:
\begin{equation}
I^{jk} =\frac12\int \rho x^j x^k dV,
\end{equation}
the first term of (\ref{b11}) can be written in terms of
partial time derivatives of $I^{ij}$ and for the static and
stationary systems it is zero. The second term represents the energy-
momentum term which can be written as~\cite{bin}:
\begin{equation}
\int\frac{\partial}{\partial x^i}(\rho\overline{u^iu_j})x^k dV = -
\int\rho \overline{u_ju^k}dV = -2K_j{}^k. \label{kin}
\end{equation}
For simplicity, we can split the kinetic tensor into the energy-momentum
tensor from the bulk velocity and the internal energy from the
dispersion velocity, yielding the result:
\begin{equation}
K_j{}^k = T_j{}^k + \frac{1}{2}\Pi_j{}^k, 
\end{equation}
where
\begin{equation}
T_j{}^k = \frac{1}{2}\int\rho \overline{u_j}~\overline{u^k} dV
\end{equation}
and where
\begin{equation}
\Pi_j{}^k = \int\rho \sigma_j{}^k dV,
\end{equation}
is the dispersion tensor with $\sigma_j{}^k =\overline{\delta u_j \delta u^k}$. The integrations of the third, fourth and fifth terms in Eq.
(\ref{b11}) contain kinetic energy contributions. The Newtonian gravitational part 
of the integrations can be interpreted as the gravitational
potential of the system. We assign the last three terms of Eq.
(\ref{b11}) to the potential energy tensor:
\begin{equation}
W_j{}^k = -\int( \delta_{ij}\Gamma^i{}_{\mu\nu}\overline{u^\mu u^\nu} + 2 \Gamma^i_{i\mu} \overline{u^\mu u_j} - \kappa\omega
\delta_{ij}B^i{}_{\mu}\overline{u^\mu})\rho x^k dV. \label{w0}
\end{equation}
Combining all the integration terms, the conservation of the energy-momentum fluid can be written as 
\begin{equation}
\frac{1}{2}\frac{d^2I_i{}^j}{d\tau^2} = 2 T_i{}^j + \Pi_i{}^j +
W_i{}^j. \label{v1}
\end{equation}
For astrophysical systems such as globular clusters,
elliptical galaxies and clusters of galaxies, the morphology of the
systems do not change in time, in other words, these systems have been 
virialized and do not interact with other systems. We set the left-hand side of 
(\ref{v1}) to zero for this case.  By choosing the center-of-mass of the system as the 
center of the coordinate system, the global velocity cancels and $T^i{}_j = 0$, whereby the energy-momentum equation simplifies to
the so-called virial equation:
\begin{equation}
\Pi_i{}^j + W_i{}^j = 0. \label{vir}
\end{equation}
In the next section, we will use the Boltzmann equation in MOG for studying clusters of galaxies. While here we derived a generic Euler equation and virial condition in MOG, we will use just the equilibrium condition for studying the clusters.  

\subsection{Equilibrium condition for Spherically Symmetric Systems: Weak Field Approximation}

In this section, we study the Boltzmann equation for spherically
symmetric systems in the weak field approximation. In order to calculate the Euler equation for a spherically symmetric
system, we start with Eq. (\ref{b11}). Let us choose $j = r$
and assume that $f$ is a function of radial distance $r$, radial
particle velocity $u_r$ and tangential particle velocity, $u_t =
\sqrt{u^2_\theta +u^2_\phi}$. For a virialized system, the shape of
the system does not change in time and terms with partial
derivatives with respect to time vanish. Then the Jeans
equation simplifies to
\begin{equation}
\frac{\partial}{\partial r}(\rho\overline{u^r u^r}) +
\Gamma^r{}_{\mu\nu}\rho\overline{u^\mu u^\nu} + 2\Gamma^i_{i r}\rho
\sigma_r^2 - \kappa \omega\delta_{ri} B^i{}_{0}\rho{u^0}= 0.
\label{b1}
\end{equation}
By assuming a spherically symmetric metric for this structure,
\begin{equation}
ds^2 =  B(r) dt^2 - A(r) dr^2 - r^2(d\theta^2 +
\sin\theta^2d\phi^2), \label{sph}
\end{equation}
we can choose $B(r) = A^{-1}(r)$ and assuming
a perturbation $B(r) = 1 + h_{00}$ and the definition 
$B^r{}_{0} = \partial^r\phi_0- \partial_0\phi^r$, the Jeans equation
in the weak field approximation can be written as
\begin{equation}
{\partial_r}(\rho \sigma_r^2) + \frac{\rho}{r}\left[ 2\sigma^2_r
-(\sigma^2_\phi+\sigma^2_\theta)\right] =
-\rho\partial_r\phi_{eff}(r). \label{jeans2}
\end{equation}

For an isotropic distribution $\rho = \rho(r)$ due to rotational
symmetry there is no preferred transverse direction i.e.,
$\sigma^2_\phi = \sigma^2_\theta$. We can still have a degree of
anisotropy in the velocity distribution at each point given by $\beta =
1- \sigma^2_\theta/\sigma^2_r$, and then the Jeans equation can be written
as
\begin{equation}
\frac{\sigma_r^2}{r} \left[\frac{d\ln\rho(r)}{d\ln r} + \frac{d\ln\sigma^2_r}{d\ln r} + 2\beta (r) \right]=  g(r),
\label{vir}
\end{equation}
where the acceleration at the right hand side, using $g(r) = - \nabla\Phi_{eff}(r)$, is given by 
$${g}({\bf{r}}) = - G_{\rm N}
\int\frac{\rho({\bf r'})({\bf r}-{\bf r'})}{|{\bf r}-{\bf r'}|^3}[1+\alpha 
-\alpha {\rm e}^{-\mu|{\bf r}-{\bf r'}|}(1+\mu|{\bf r}-{\bf r'}|) ]d^3{\bf r'},$$ 
where for the scales $|r - r'|\ll \mu^{-1}$, the exponential term approaches 
to one and by canceling the last two terms in the integrand, the gravitational 
acceleration reduces to Newtonian gravity. On the other hand, for 
the case that $|r - r'|\gg \mu^{-1}$, the exponential term approaches to 
zero and we get Newtonian gravity with an enhanced gravitational constant 
$(1 + \alpha)G_N$. In what follows, we can set $\beta = 0$ in Eq. (\ref{vir}) for the isotropic 
distribution of the velocity. 

\begin{table*}
\caption[]{Sample of 11 clusters of galaxies observed by the Chandra telescope. We adapt this table from~\cite{chandra}. The first column is the name of the cluster, the second column is the radius of the cluster with the cluster density $500$ times larger than the background. The other columns give the parameters of the density profile in Eq. (\ref{modne}).}
\begin{tabular}{|c|c|c|c|c|c|c|c|c|c|c|}
\hline\hline
Cluster &
$r_{500}$ &
{$n_0$} &
{$r_c$} &
{$r_s$} &
{$\alpha^\prime$} &
{$\beta^\prime$} &
{$\varepsilon$} &
{$n_{02}$} &
{$r_{c2}$} &
{$\beta_2$}\\

 & {(kpc)} & {$10^{-3}$~cm$^{-3}$}
 & {(kpc)} & {(kpc)} & & & & {$10^{-1}$~cm$^{-3}$} & & \\
\hline
A133          \dotfill & $1007\pm41$&  4.705 & 94.6 & 1239.9 & 0.916 & 0.526 & 4.943 &   0.247 &    75.83 &   3.607     \\
A262          \dotfill &  $650\pm21$&  2.278 &  70.7 &  365.6 & 1.712 & 0.345 & 1.760 & -- &    -- & --      \\
A383          \dotfill &  $944\pm32$&   7.226 & 112.1 &  408.7 & 2.013 & 0.577 & 0.767 &   0.002 &     11.54 &   1.000 \\
A478          \dotfill & $1337\pm58$&   10.170 & 155.5 & 2928.9 & 1.254 & 0.704 & 5.000 &   0.762 &      23.84 &   1.000     \\
A907          \dotfill & $1096\pm30$ &   6.257 & 136.9 & 1887.1 & 1.554 & 0.594 & 4.986 & -- &    -- & -- \\
A1413         \dotfill & $1299\pm43$&   5.239 & 195.0 & 2153.7 & 1.247 & 0.661 & 5.000 & -- &    -- & -- \\
A1795         \dotfill & $1235\pm36$&    31.175 &  38.2 & 682.5 & 0.195 & 0.491 & 2.606 & 5.695 &  3.00 & 1.000     \\
A1991         \dotfill & $732\pm 33$&   6.405 &  59.9 &  1064.7 & 1.735 & 0.515 & 5.000 &   0.007 &     5.00 &   0.517     \\
A2029         \dotfill & $1362\pm43$&    15.721 &  84.2 &  908.9 & 1.164 & 0.545 & 1.669 &   3.510 &     5.00 &   1.000     \\
A2390         \dotfill & $1416\pm48$&   3.605 & 308.2 & 1200.0 & 1.891 & 0.658 & 0.563 & -- &    -- & -- \\
MKW4          \dotfill &  $634\pm28$&   0.196 & 578.5 & 595.1 & 1.895 & 1.119 & 1.602 &   0.108 &      30.11 &   1.971  \\
\hline
\end{tabular}
\label{table1}
\end{table*}

\section{X-ray Chandra cluster data}
\label{data}

We now apply Eq. (\ref{vir}) for the spherically symmetric X-ray clusters of galaxies. For these systems, we have two observables: (i) the ionized gas profile $\rho_g(r)$ and (ii) the temperature profile $T(r)$ where both functions are determined from the center of a cluster. We can relate the temperature and gas profile to the dispersion velocity of gas and Eq. (\ref{vir}) can be written as~\cite{sarazin}:
\begin{equation}
\frac{k_B T(r)}{\mu_p m_p r}\left(\frac{d\ln\rho_g(r)}{d\ln r} + \frac{d\ln T(r)}{d\ln
r}\right) = g(r), \label{halo}
\end{equation}
where $k_B T(r) = \mu_p m_p \sigma_r^2$, $k_B$ is Boltzmann's constant, $m_p$ denotes the mass of the proton and $\mu_p = 0.6$ represents the average value for baryonic mass. The left-hand side of equation (\ref{halo}) 
is determined by observational data, which has to be consistent 
with the acceleration determined on the right-hand side by the distribution of matter. In Newtonian gravity without 
dark matter, the left-hand side is one order of magnitude larger than the right-hand side and in order to have a consistent equation, 
dark matter needs to be added to the gravitating component of the cluster~\cite{chandra}.

In the following, we use the gas density and temperature profiles of clusters and compute the left-hand side of equation (\ref{halo}) from the observational data~\cite{chandra}.

\subsection{Gas-density profile}

In clusters of galaxies the temperature of the gas is of order keV. For this temperature the gas is fully ionized and the hot plasma is mainly emitted by the free-free radiation (bremsstrahlung) process. There is also line emission by the ionized heavy elements. The radiation process generated by these two mechanisms is proportional to $n_e n_p$, namely, the number density of free electrons times the number density of protons.

The standard result for the distribution of a gas plasma in a cluster of galaxies is called the $\beta$-model. The three-dimensional density is given by~\cite{cav}:
\begin{equation}
n_e n_p = \frac{n_0^2}{(1+r^2/r_c^2)^{3\beta^\prime}}.
\label{nee}
\end{equation}
We will use the density profile for the sample of clusters by modifying equation (\ref{nee}) in such a way that (i) it has a cusp at the center, (ii) at large radii X-ray brightness is steeper than the $\beta$-model, (iii) adding the second $\beta$-model component with smaller core which produces more freedom near the center of the cluster \cite{chandra}. We obtain
\begin{eqnarray}
n_en_p &=&  \frac{ (r/r_c)^{-\alpha^\prime}}{(1+r^2/r_c^2)^{3\beta^\prime-\alpha^\prime/2}}\frac{n_0^2}{(1+r^\gamma/r_s^\gamma)^{\epsilon/\gamma}}\nonumber \\
&+&\frac{n_{02}^2}{(1+r^2/r_{c2}^2)^{3\beta_2}}.
\label{modne}
\end{eqnarray}

By considering the primordial abundance of He and the relative metallicity compared to the sun, $Z = 0.2 Z_\odot$, we obtain for the baryonic density of the cosmic plasma gas:
\begin{equation}
\rho_g(r) = 1.24 m_p \sqrt{n_e(r) n_p(r)}.
\label{density}
\end{equation}
Table (\ref{table1}) represents the best fit to the density function in Eq. (\ref{modne}) for 11 clusters of galaxies in the Chandra catalog~\cite{chandra}.

For the stellar component of clusters, we use the empirical relation between the stellar and the Newtonian mass~\cite{vikh2,vikh3}:
\begin{equation}
\frac{M_{\star}}{10^{12}M_{\odot}} = (1.8\pm0.1) \biggl(\frac{M_{500}}{10^{14}M_{\odot}}\biggr)^{0.71\pm0.04},
\label{stars}
\end{equation}
where $M_{500}$ is the mass of a cluster in the dark matter scenario.

\subsection{Temperature profile}

The observed temperature profile is similar to the density profile in a two-dimensional projection of a physical three-dimensional distribution of temperature~\cite{chandra}. Since the plasma in the cluster can be divided into two regions, the cooling zone at the center and the outer part of the cluster, the temperature profile for these two regions can be given by two different functions. For the outside of the central cooling zone, the temperature is given by:
\begin{equation}
T_{out}(r) = \frac{(r/r_t)^{-a}}{\big[1 + (\frac{r}{r_t})^b)\big]^{c/b}}.
\end{equation}
At the center of the cluster the temperature declines due to cooling in this region and it is given by~\cite{allen}:
\begin{equation}
T_{in}(r) = \frac{\biggl(x + T_{min}/T_0\biggr)}{x+1},~~~ x = \biggl(\frac{r}{r_{cool}}\biggr)^{a_{cool}}.
\end{equation}
The overall three-dimensional temperature profile of the cluster is the multiplication of 
these two temperature functions:
\begin{equation}
T_{3D}(r) = T_0 T_{in}(r) T_{out}(r).
\label{Tprofile}
\end{equation}
Table (\ref{tabel2}) represents the best fit to the temperature profile of clusters. 
\begin{table*}
\caption[]{Sample of 11 clusters of galaxies observed by the Chandra telescope~\cite{chandra}. The first column is the name of the cluster and the other columns are the parameters of the temperature profile in Eq. (\ref{Tprofile}).}
\begin{tabular}{|c|c|c|c|c|c|c|c|c|}
\hline\hline
Cluster &
$T_0$ &
{$r_t$} &
{$a$} &
{$b$} &
{$c$} &
{$T_{min}/T_0$} &
$r_{cool}$&
$a_{cool}$\\
 & {(Kev)} & {(Mpc)} & & & &  & {(kpc)} &   \\
\hline
A133          \dotfill & 3.61 &      1.42 & 0.12 &  5.00 & 10.0 & 0.27   & 57     & 3.88      \\
A262          \dotfill & 2.42 &      0.35 &  -0.02 & 5.00   & 1.1 & 0.64  & 19     & 5.25      \\
A383          \dotfill & 8.78 &      3.03 & -0.14 &  1.44 & 8.0  & 0.75   & 81     &  6.17  \\
A478          \dotfill & 11.06&     0.27 & 0.02 &  5.00 & 0.4    & 0.38  & 129    & 1.60     \\
A907          \dotfill & 10.19&     0.24 & 0.16 & 5.00 & 0.4     & 0.32  & 208    & 1.48   \\
A1413         \dotfill & 7.58 &     1.84 & 0.08 & 4.68 & 10.0   & 0.23  & 30      & 0.75   \\
A1795         \dotfill & 9.68 &     0.55 &  0.00 & 1.63 & 0.9    & 0.10  & 77      & 1.03         \\
A1991         \dotfill & 2.83 &     0.86 &  0.04 &  2.87 & 4.7   & 0.48  & 42      &  2.12       \\
A2029         \dotfill & 16.19&     3.04 &  -0.03 &  1.57 & 5.9 & 0.10  & 93      &  0.48        \\
A2390         \dotfill & 19.34&     2.46 &  -0.10 & 5.00 &10.0 & 0.12  & 214    & 0.08  \\
MKW4          \dotfill &  2.26&    0.10 & -0.07 & 5.00 & 0.5    & 0.85  & 16      &  9.62      \\
\hline
\end{tabular}
\label{tabel2}
\end{table*}
\begin{table*}
\begin{center}
\caption[]{The overall dynamical mass in MOG versus the observed baryonic mass of clusters. 
The columns describe: (1) the name of the cluster from the Chandra 
catalog~\cite{chandra}, (2) the mass of the gas from integrating over the density of the gas up to the detected radius, (3) the empirical mass of stars from equation (\ref{stars}), (4) the overall baryonic mass (the error bars are adopted from the direct observations,~\cite{chandra}) and, (5) the dynamical mass inferred from Eq. (\ref{master}). The error bars adapted from upper and lower values of $\alpha$ from \cite{rah13}.}
\begin{tabular}{|c|c|c|c|c|}
\hline\hline
Cluster (1)&
$M_{gas}$(2) & 
$M_{stars}$(3) & 
$M_{baryonic}$(4) &
$M_{dyn}$(5) \\

  & $\times 10^{13} M_\odot$  & $\times 10^{13} M_\odot$   & $\times 10^{13} M_\odot$ & $\times 10^{13} M_\odot$ \\ 
  
\hline
A133   \dotfill &$2.61\pm0.54$  & 0.408 & $3.02\pm0.54$ &$ 4.12_{-0.14}^{+0.15}$ \\
A262          \dotfill &  $0.92\pm0.16$ & 0.13 & $1.05\pm0.16$  &$1.35_{-0.05}^{+0.05}$ \\ 
A383          \dotfill & $3.72\pm 0.59$  &  0.398 & $4.12\pm 0.59$  & $4.19_{-0.14}^{+0.15}$ \\ 
A478          \dotfill & $9.06 \pm 2.04$ & 0.765  & $9.82\pm 2.04$ & $7.71_{-0.25}^{+0.28}$ \\ 
A907          \dotfill & $ 5.46\pm0.72$   & 0.528 & $5.99\pm0.72$ &  $5.09_{-0.17}^{+0.13}$ \\ 
A1413         \dotfill &$ 7.86\pm 1.34$  & 0.527  & $8.39\pm 1.34$ & $9.30_{-0.37}^{+0.33}$ \\ 
A1795         \dotfill & $6.18\pm0.90 $ &0.644  & $6.83\pm0.90$ & $7.08_{-0.23}^{+0.26}$\\ 
A1991         \dotfill &$1.25\pm0.27 $  &0.208 & $1.45\pm0.27$ & $1.57_{-0.05}^{+0.06}$ \\
A2029         \dotfill &$9.47\pm 1.46$  & 0.788& $10.26\pm 1.46$ & $10.69_{-0.35}^{+0.38}$ \\ 
A2390         \dotfill &$15.03\pm2.46 $ &0.787 &$15.82\pm2.46$ & $13.78_{-0.46}^{+0.49}$ \\ 
MKW4          \dotfill &$0.47\pm0.08$ & 0.149 &$0.62\pm0.08$ & $0.76_{-0.03}^{+0.03}$ \\


\hline
\end{tabular}
\label{tab3}
\end{center}
\end{table*}

\subsection{Virial Theorem and X-ray Clusters}
\begin{figure}
\begin{center}
\includegraphics[width=70mm]{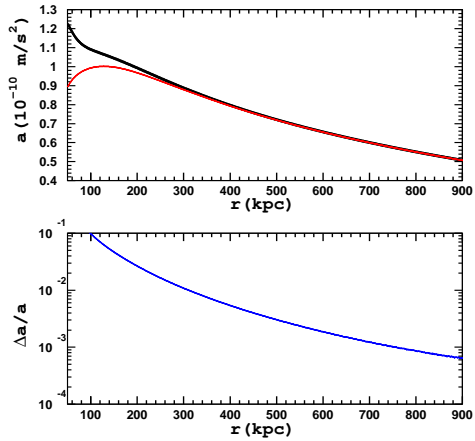}
\caption {Upper panel: Comparing the acceleration of a test mass particle in the cluster of galaxies as a function of distance from the center of the cluster, using equation (\ref{potential22}) for two cases of numerical integration taking into account the attractive and repulsive forces (red thin line) and 
ignoring the repulsive force at the large scales (thick black line). The lower panel represents the relative 
difference between the two calculations.\label{acc}} 
\end{center}
\end{figure}

We use the observed density of gas and temperature profiles in Eq. (\ref{density}) and (\ref{Tprofile}) with the corresponding parameters for each cluster to calculate the left-hand side of equation (\ref{halo}). On the other hand, on the right-hand side of this equation, we can use the baryonic matter of a cluster composed of plasma gas and galaxies to calculate the acceleration of a test mass particle. For a cluster or any complicated structure, the acceleration of a test particle with the MOG effective potential in Eq. (\ref{potential22}) can be obtained by numerical integration.

\setcounter{figure}{1}
\begin{figure*}
\begin{center}
\begin{tabular}{cc}
\includegraphics[width=70mm]{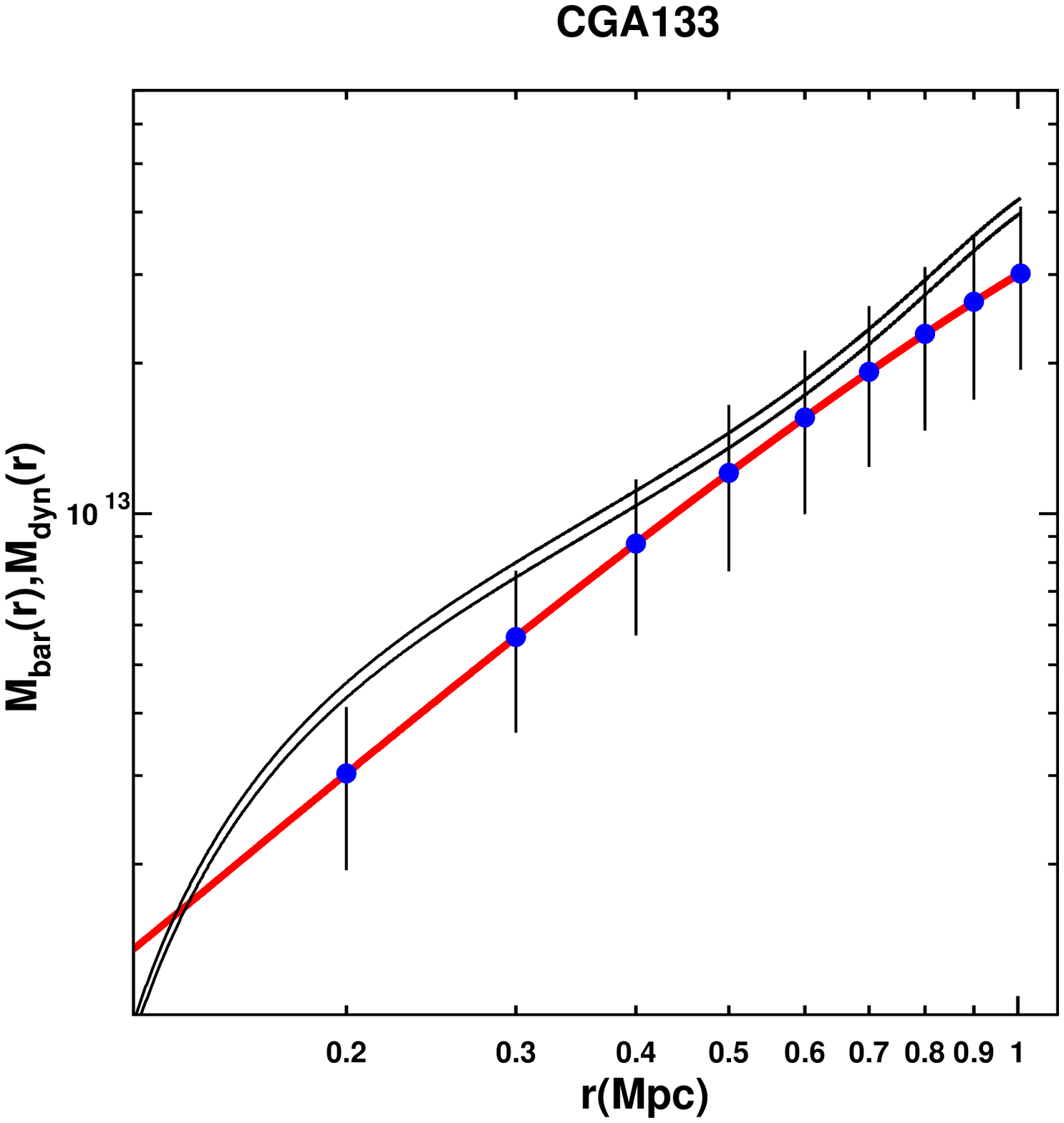}&
\includegraphics[width=70mm]{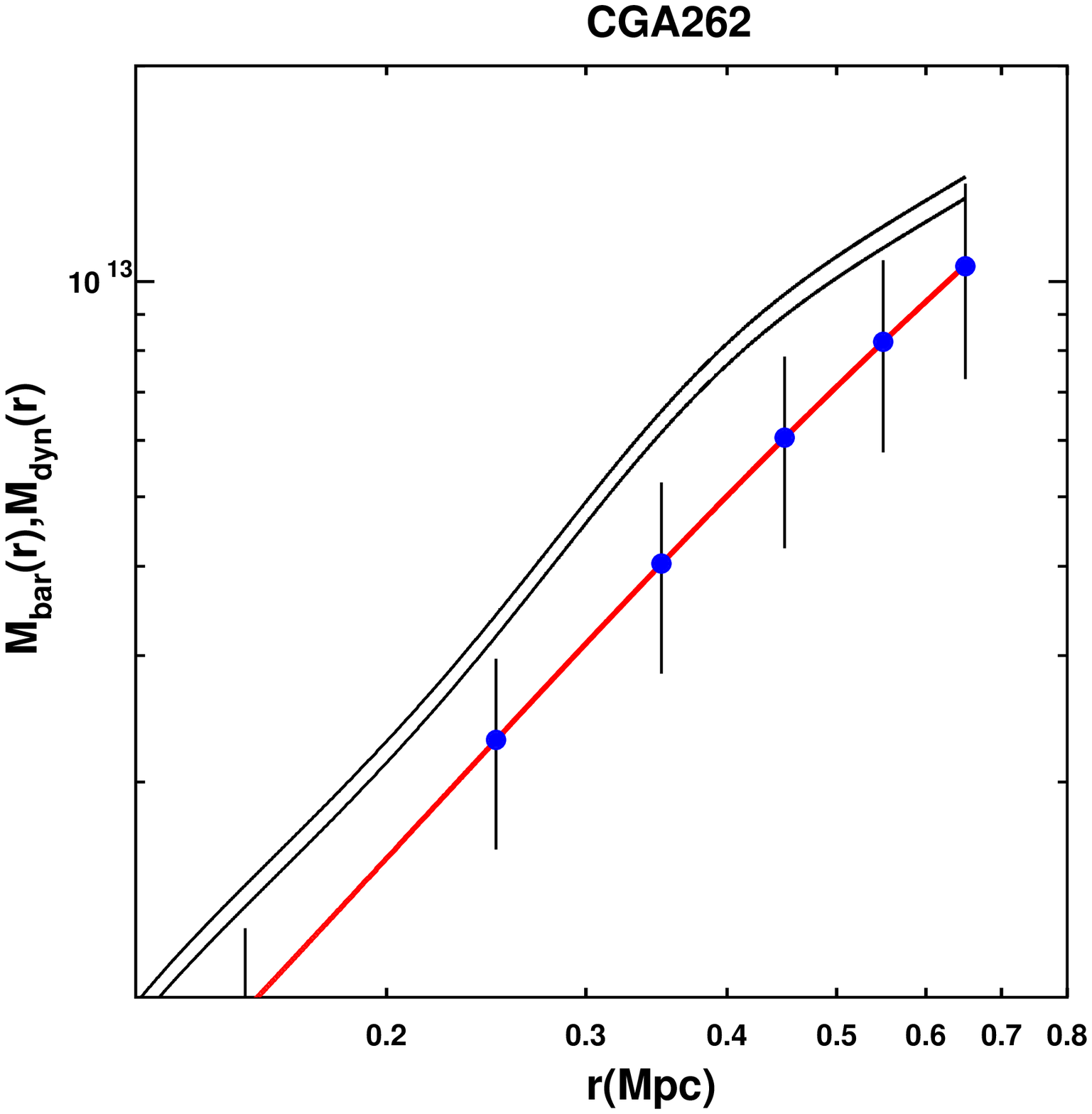} \\
\includegraphics[width=70mm]{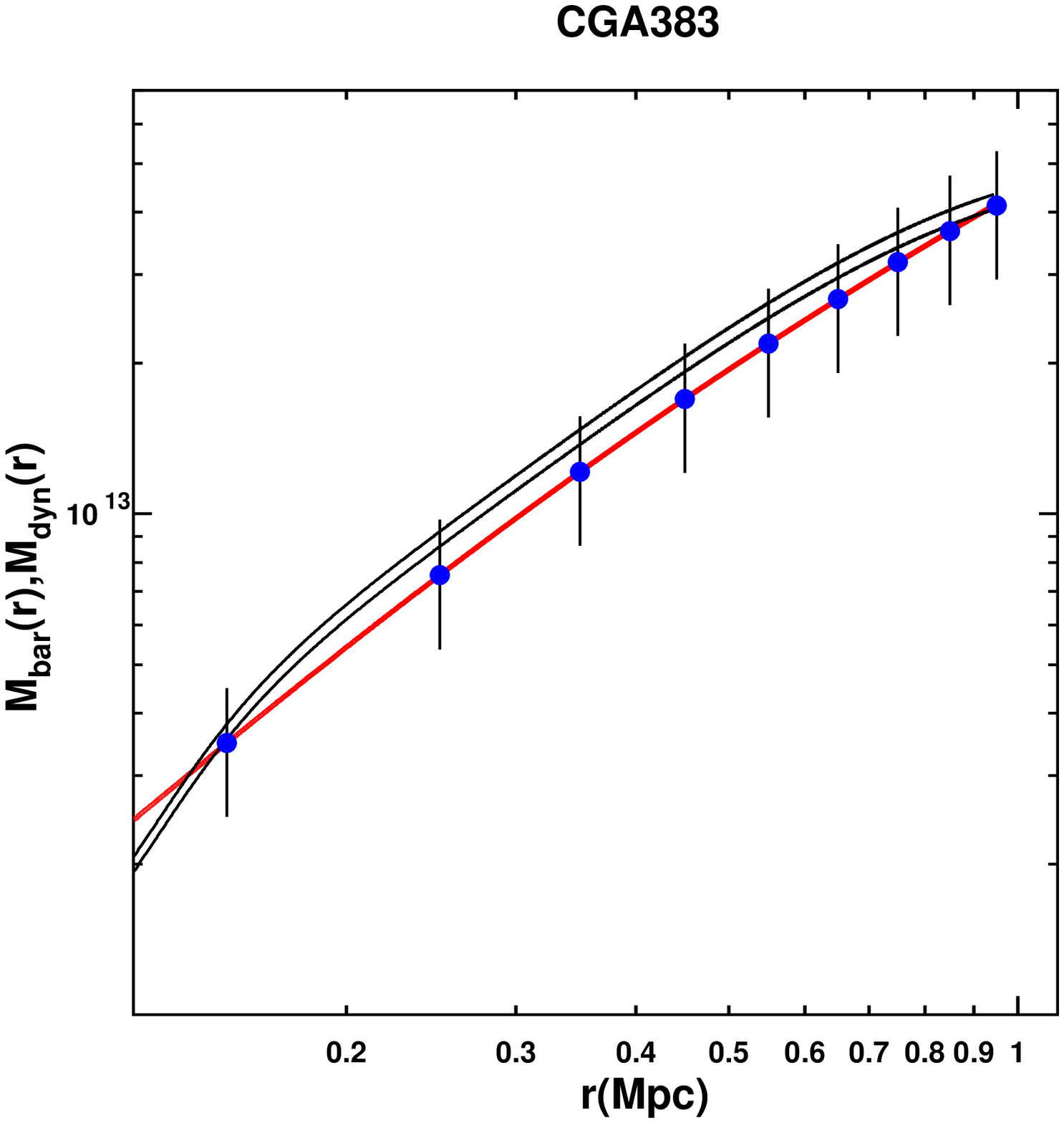} &
\includegraphics[width=70mm]{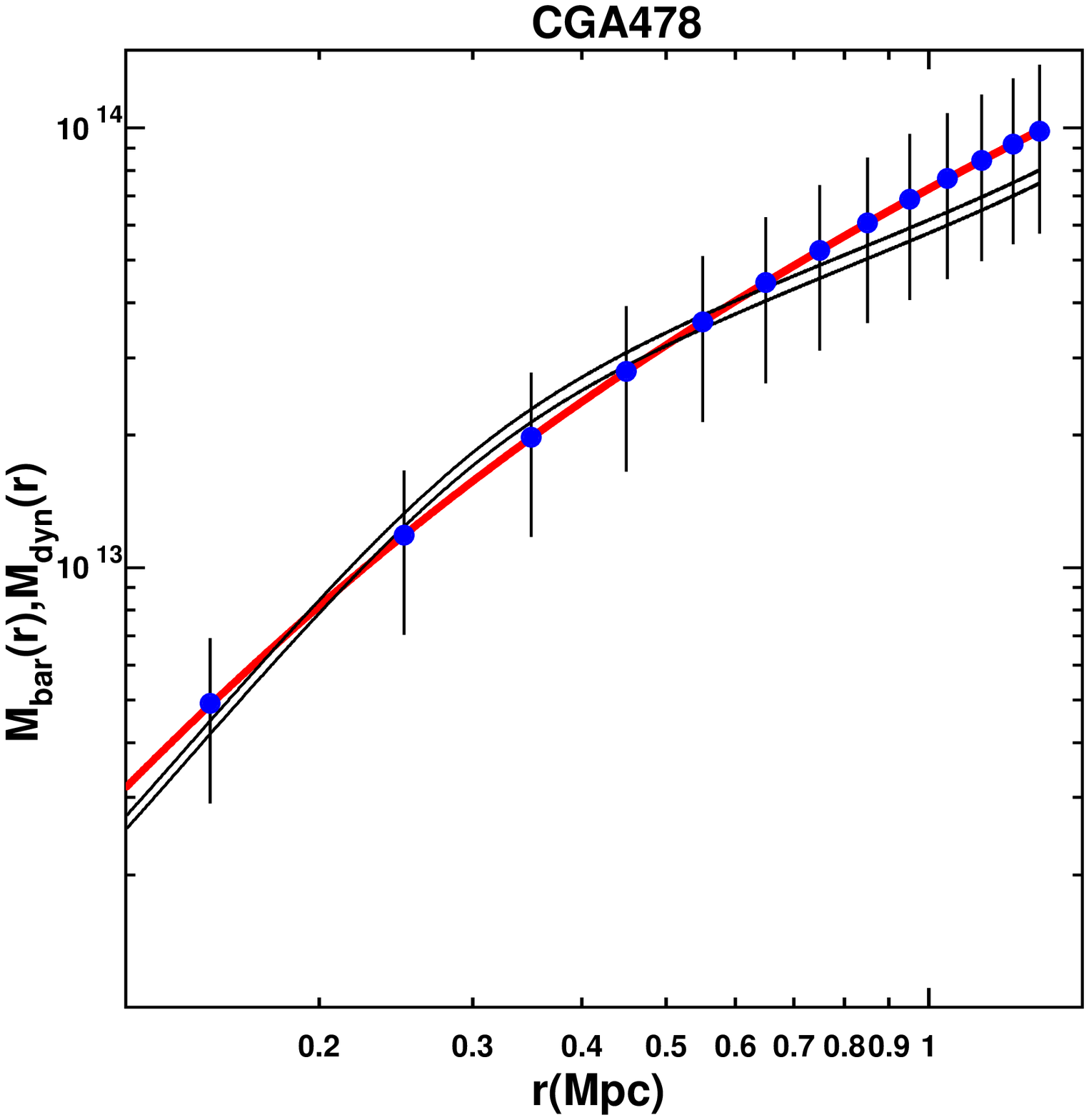}\\
\includegraphics[width=70mm]{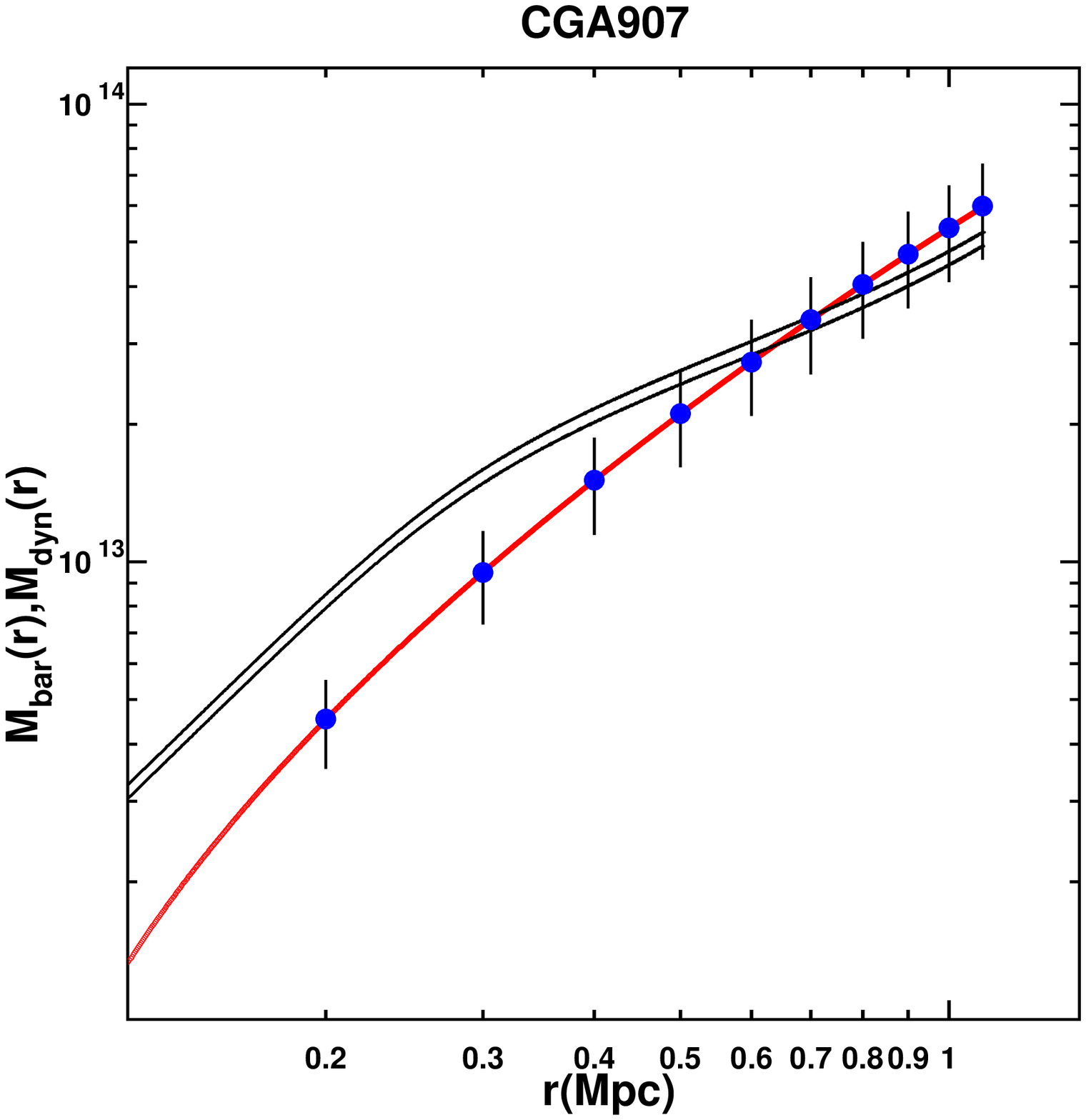}&
\includegraphics[width=70mm]{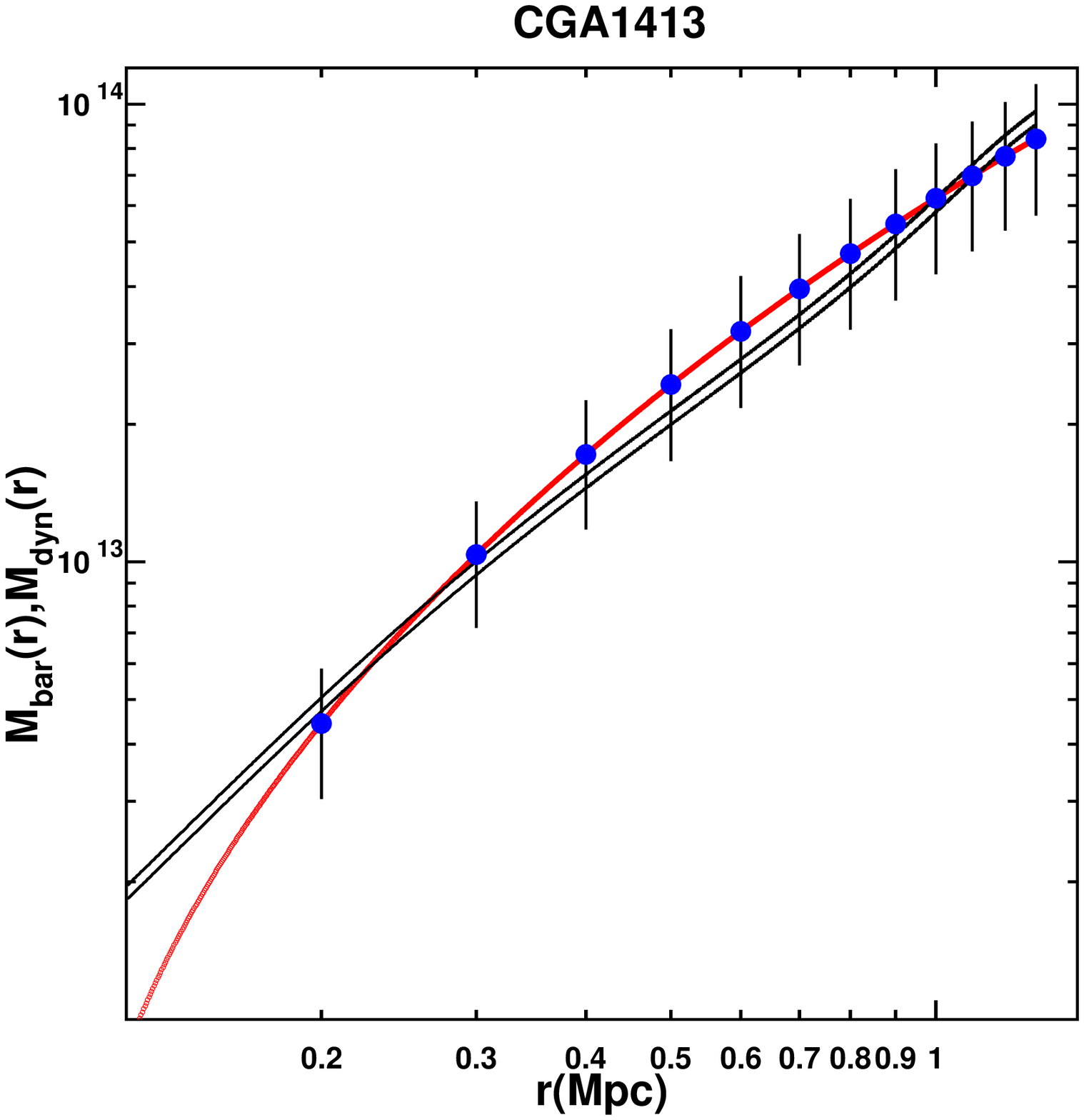}\\
\end{tabular}
\end{center}
\caption {The black solid lines represent the dynamical mass 
in MOG with the uncertainty resulting from the error bar in the value of $\alpha = 8.89 \pm 0.34$~\citep{rah13}. The red line represents the baryonic mass including the mass of the gas and the mass of the stars in the cluster. The error bars represent uncertainty in the measurements of the mass of the clusters \citep{chandra}. 
\label{fig2} }
\newpage
\end{figure*}

\setcounter{figure}{1}
\begin{figure*}
\begin{center}
\begin{tabular}{cc}
\includegraphics[width=70mm]{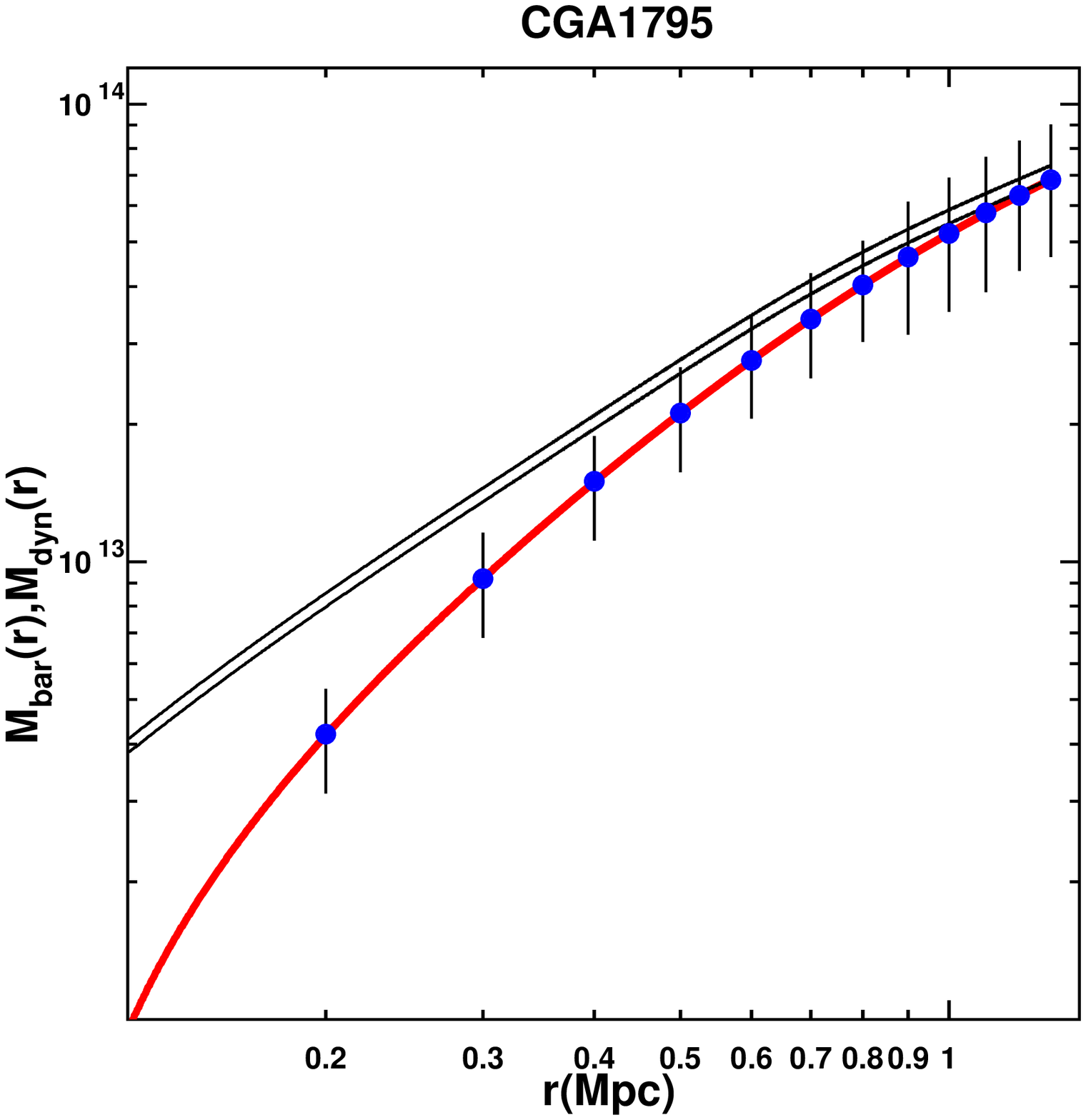}&
\includegraphics[width=70mm]{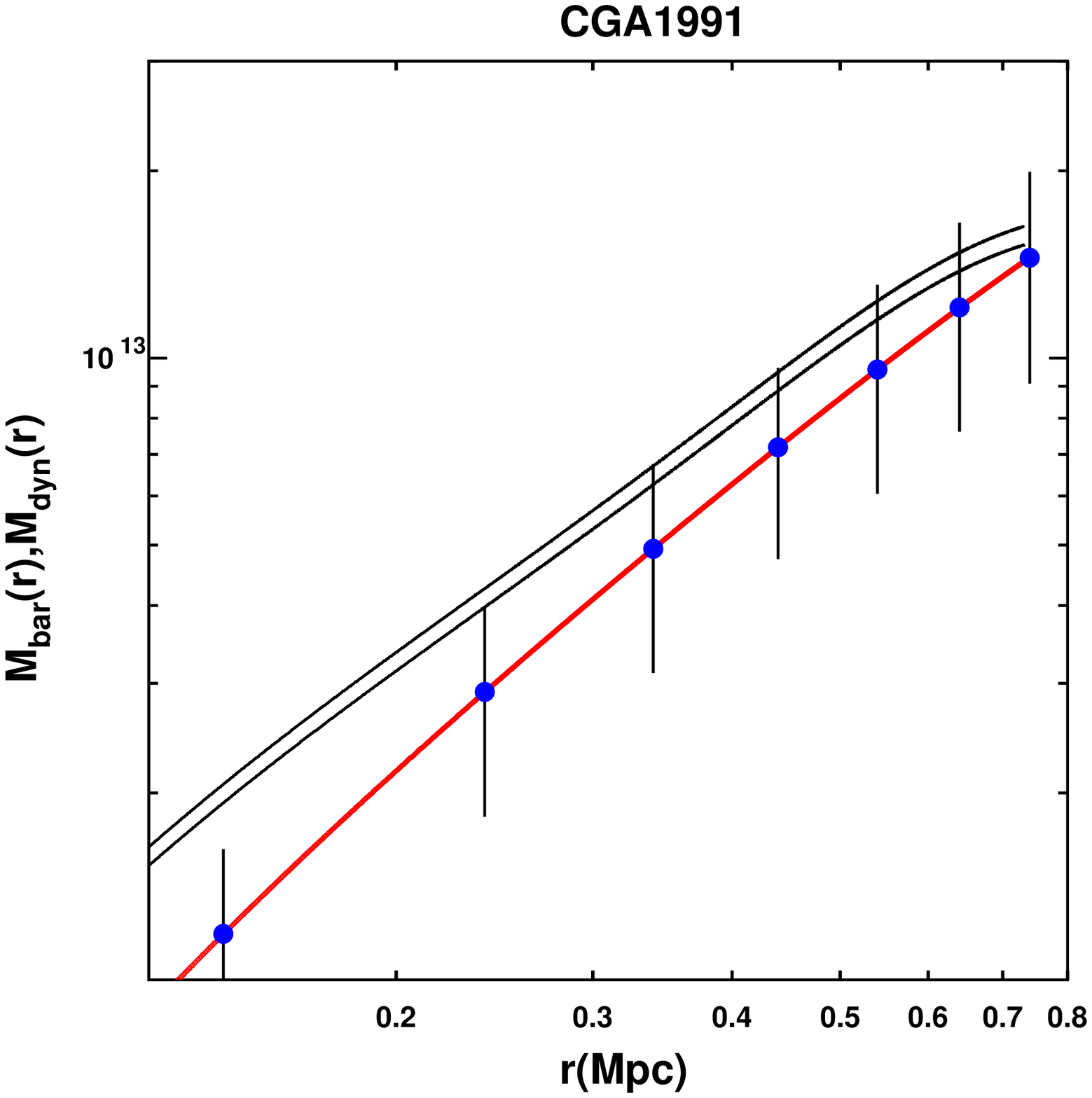}\\
\includegraphics[width=70mm]{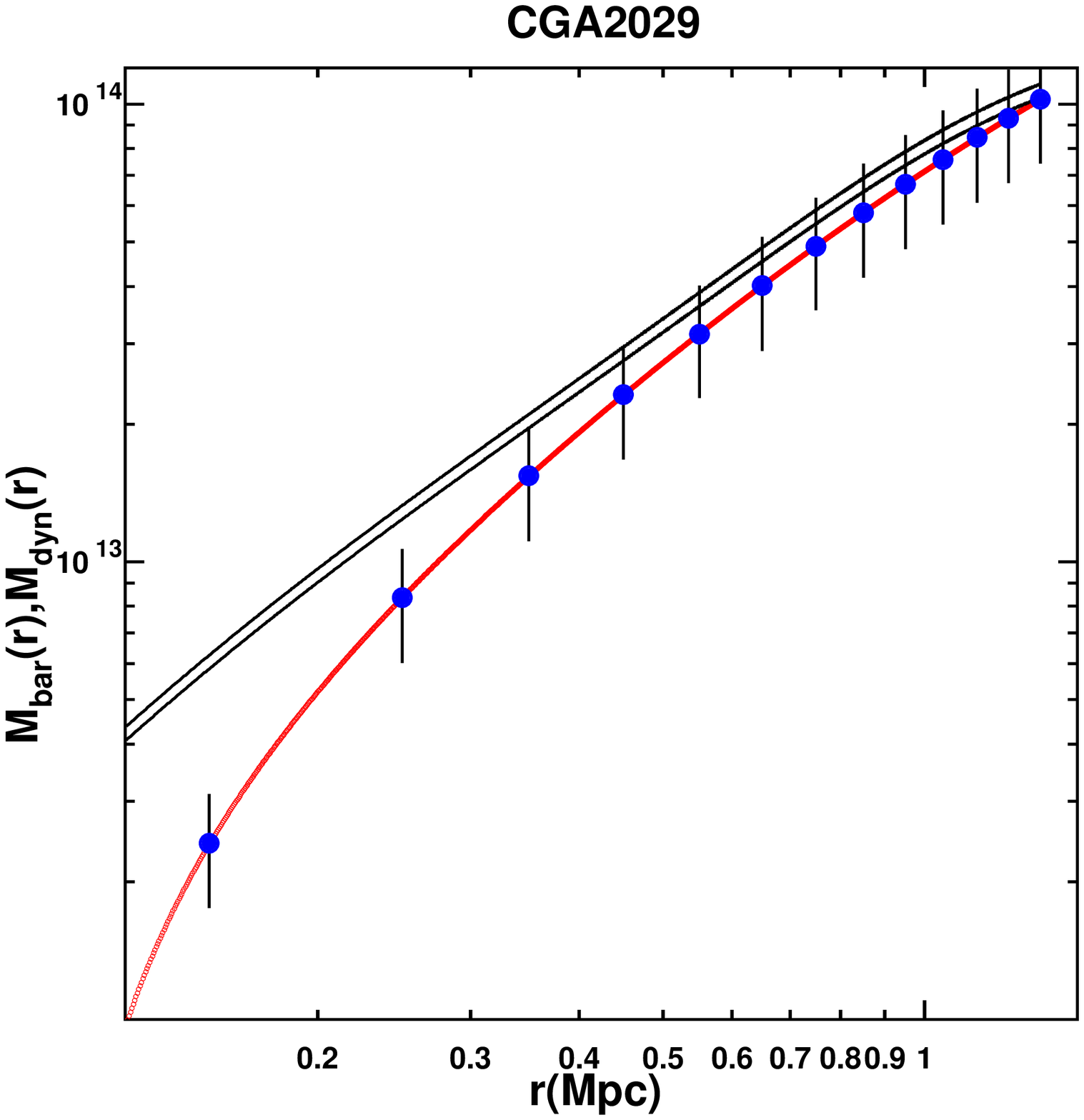}&
\includegraphics[width=70mm]{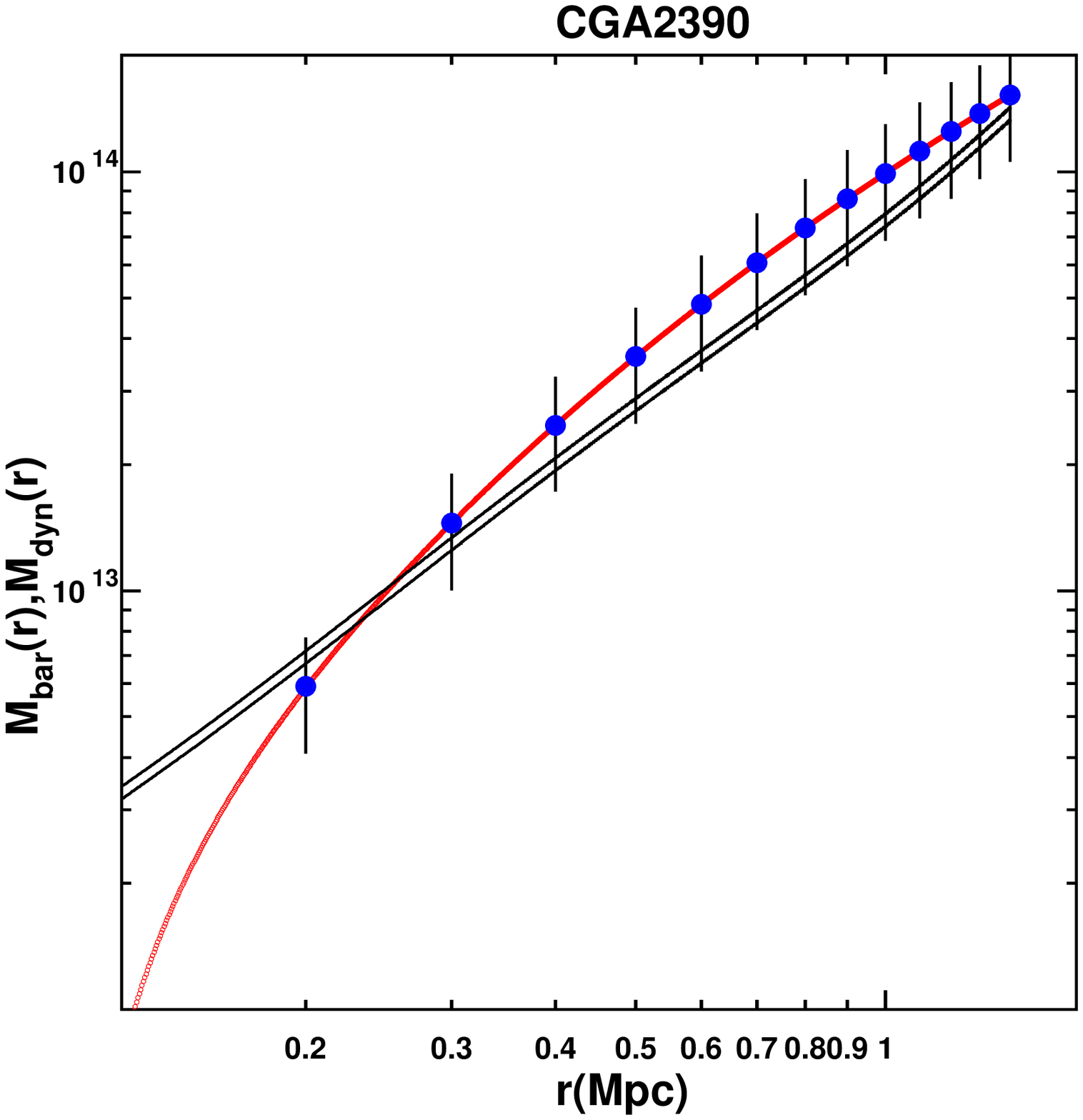}\\
\includegraphics[width=70mm]{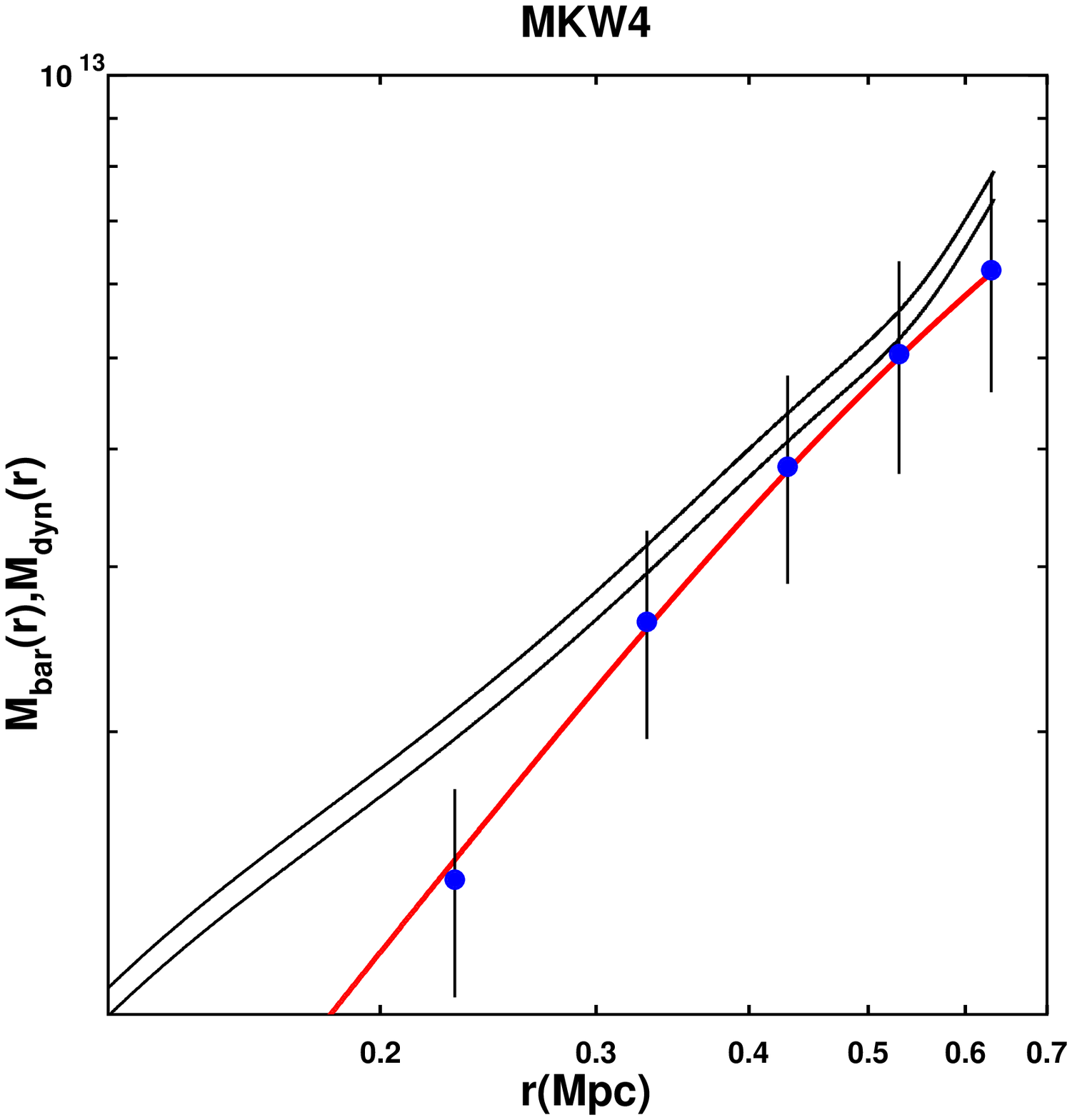}&
\end{tabular}
\end{center}
\caption { .... continued
\label{fig2} }
\newpage
\end{figure*}

\begin{figure*}
\begin{center}
\includegraphics[width=70mm]{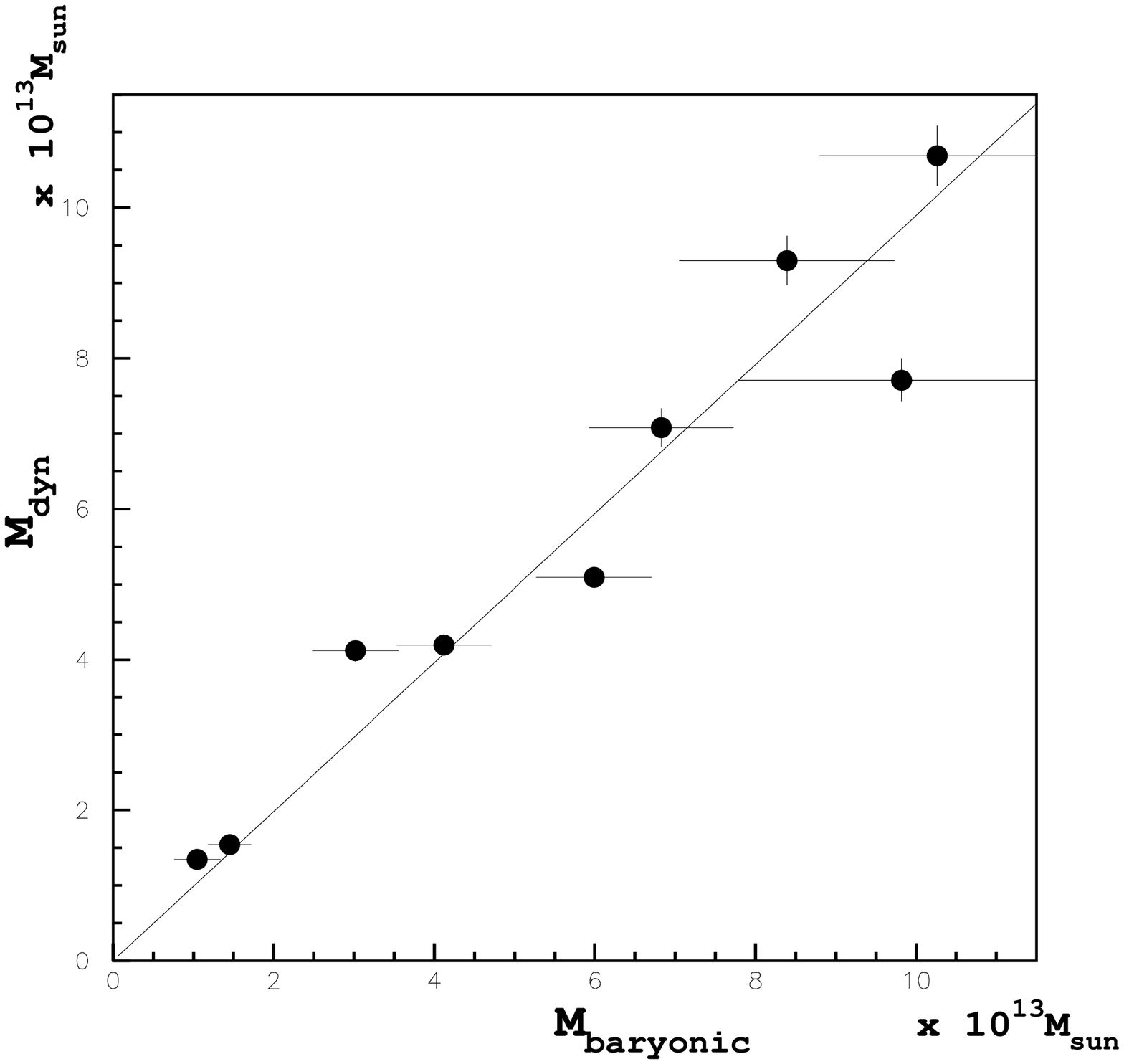} 
\includegraphics[width=70mm]{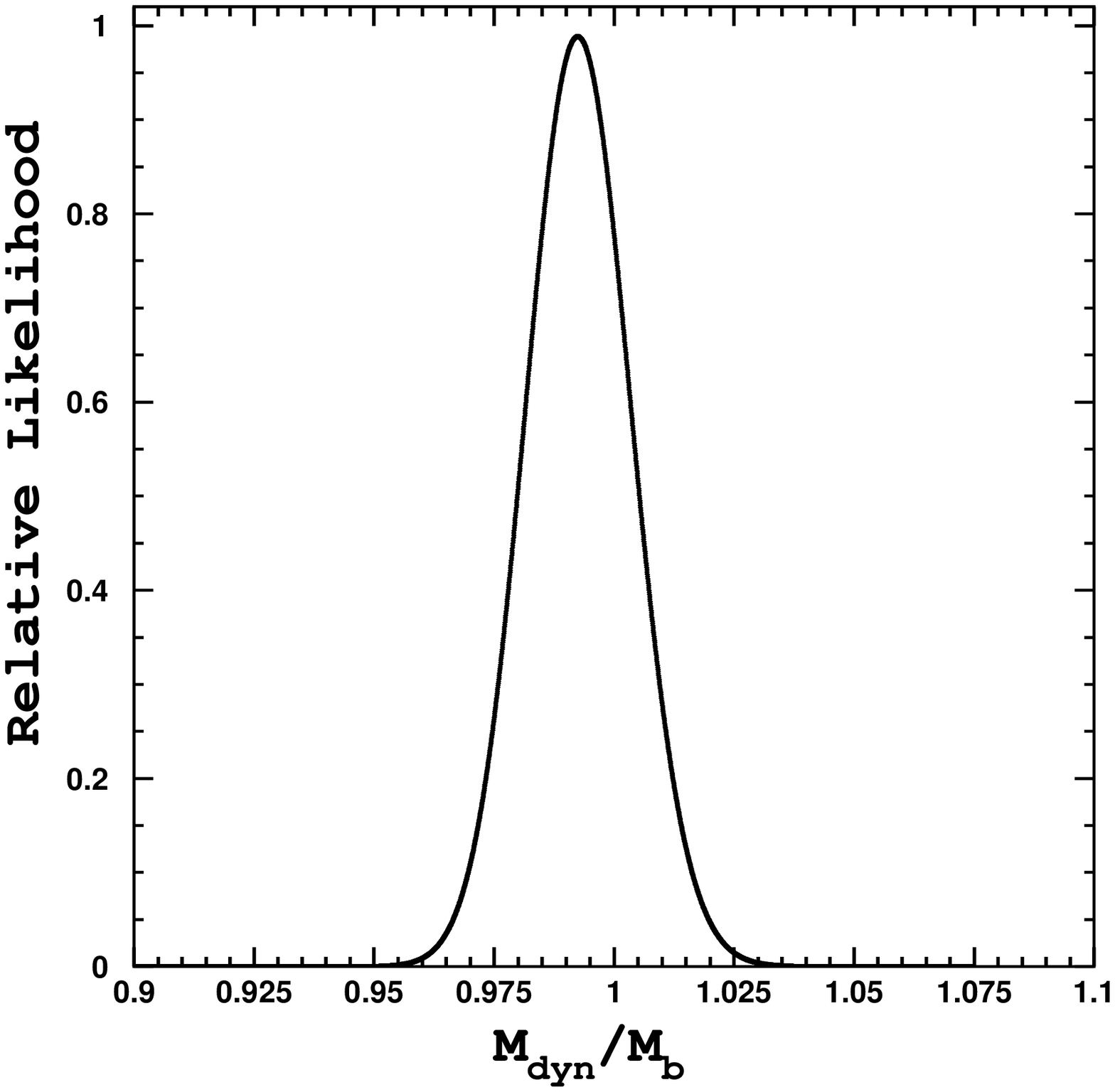}
\caption {Comparison of the dynamical mass in MOG versus the baryonic mass for the sample of clusters
in Table (\ref{tab3}). The baryonic mass is composed of gas and stars. The filled circles indicate the corresponding masses up to $r_{500}$ with the corresponding error bars. The solid line shows the best fit to the linear relation $M_{\rm dyn} = \beta_{\rm cl} M_{\rm bar}$ between the two masses with the best value of $\beta = 0.99$. The likelihood function for this fit is given at the right panel.
\label{fig3}
} 
\end{center}
\end{figure*}

Here, for a spherical symmetric structure with scales of the order of $\ell_{\rm cl}\sim 1$ Mpc, we
can simplify our numerical calculation, noting that the length corresponding to the mass of the 
vector field, $\mu^{-1}\simeq 24$ kpc, is much smaller than the size of a cluster. For these structures, the gradient of the density inside the cluster within the length scale of $\mu^{-1}\simeq 24$ kpc is negligible and we can ignore the gravitational acceleration exerted on a test particle by the fluid surrounding it within this radius. Hence, the second term of Eq. (\ref{potential22}) for a cluster-size structure is negligible. We have numerically checked this simplification by comparing the overall effective gravitational acceleration with the simplified relation of keeping only the first term of equation (\ref{potential22}). Figure (\ref{acc}) compares the relative acceleration on a test mass particle with and without considering the Yukawa term. Since 
the measurements of central parts of clusters in the Chandra observations are not accurate enough and the data is scattered~\cite{priv}, using the simplified formula for the acceleration at the outer radii of the clusters (i.e. $r>100$ kpc) and ignoring the Yukawa term is considered to be valid.

Using Gauss's theorem for a spherically symmetric system with a $1/r$ potential, the dynamical mass from 
Eq. (\ref{halo}) can be written as 
\begin{equation}
M_{\rm dyn}(r) = - 3.68\times 10^{10}\frac{r  T(r)} {1+\alpha}  \left(\frac{d\ln \rho_g(r)}{d\ln r}  +\frac{d\ln T(r)}{d\ln r} \right ),
\label{master}
\end{equation}
where $M_{\rm dyn}$ is given in terms of solar mass, $r$ is the distance from the center of the cluster in kpc, $T(r)$ is the temperature of the gas in keV.  We use the value $\alpha = 8.89 \pm 0.34$ which has been universally fixed by the fits to the rotation curve data of galaxies~\cite{rah13}. 

In Figure (\ref{fig2}), we compare the profile of dynamical mass $M_{dyn}(r)$ as a function of distance from the center of the cluster with the baryonic mass of clusters composed of the mass of gas plus the mass contribution from stars, $M_{bar}(r) = M_{gas}(r) + M_\star(r)$. The overall dynamical and baryonic masses at the distance $r_{500}$ are given in Table (\ref{tab3}). The error bar associated to the mass of baryonic component is adapted from Vikhlinin et al (2006) where they reported the fraction of baryonic mass 
to the overall mass in the Dark matter scenario. Assuming the error bar corresponds to this fractional mass as a constant value throughout the cluster, we calculate the error bar of the baryonic mass being proportional to the mass. We plot the error bars for the baryonic matter of clusters as a function of distance from the center of the cluster and with the cadence of $0.1$ Mpc separation. 

For some of clusters as CGA262 and CGA1991 there is an offset between the observation and theoretical 
mass profiles with about $1\sigma$ deviation.  For the rest of the clusters there is consistency between the theoretical and observational mass profiles, except for the central regions of clusters. We note that observational data at the central parts of clusters is poor.  Moreover, this part might not completely virialized due to cooling flow. For most of the clusters at scales larger than $300$ kpc, the 
theoretical profiles are in good agreement with the observational data. From the mass profiles in Figure (\ref{fig2}), the 
outer regions of clusters at $r>300$ kpc contain almost $90\%$ of the mass of the clusters and for these regions the MOG theory prediction follows the observational data. For the overall mass of the clusters at $r_{500}$, we compare the mass profiles for the data and theory. The error bar associated with the observed mass is obtained from the uncertainty in the observational data, while for the theoretical profile, we obtain the upper and lower values for the mass resulting from the uncertainties in $\mu$ and $\alpha$, reported in~\cite{rah13}. Figure (\ref{fig3}) compares the overall theoretical baryonic mass of clusters with the observational data. 

\section{Conclusions}
\label{conc}

Following our previous paper~\cite{rah13}, in this work we have investigated further the weak field approximation 
of STVG to test the dynamics of clusters of galaxies. The theory of gravity in the weak field regime is composed of an enhanced Newtonian gravity term with the gravitational constant $G \simeq 10 G_N$ and a Yukawa repulsive term with the mass $\mu^{-1} \simeq 24$ kpc.  At the smaller scales compared to $\mu^{-1}$, e.g., solar system scales, standard Newtonian gravity can be recovered. On the other hand, at larger scales we get Newtonian gravity with a stronger effective gravitational constant. Our analysis of a sample of galaxies in the previous work showed that with the fixed universal parameters $\alpha$ and $\mu$ in the effective gravitational potential~\cite{rah13}, we can successfully describe the dynamics of a large set of galaxies with only their baryonic matter content.

In order to test this universality, we have analyzed a sample of nearby Chandra X-ray data in which the gas density profiles and the temperatures have been measured~\cite{chandra}. Our analysis shows that, using the universal values for the parameters $\alpha$ and $\mu$ in the effective MOG potential, we get consistent mass profiles and overall values for the dynamical and baryonic mass of clusters, except for the central parts of clusters where the observational data is scattered.  These results along with the analyses of the rotation curves of galaxies demonstrate that with a unique weak field MOG effective potential, we can describe the dynamics of astrophysical systems from the solar system to $\sim $Mpc scale systems without exotic dark matter. 

Finally, we outline further questions in MOG yet to be answered, (i) large scale structure formation in the universe. The structure formation growth has to be described by MOG without using dark a matter component and using CMB boundary conditions, (ii) interpretation of observational data of gravitational lensing at large scales, especially the case of the asymmetric Bullet Cluster without invoking dark matter for the lensing system. A previous published investigation of the Bullet Cluster using  a version of MOG~\cite{b2,b3}, showed compatibility between the predictions of MOG and the available data at the time. However, this work was based on a spherically symmetric point-like solution of the field equations, which has to be extended to continuous matter distributions using the methods of the present paper.

\section*{Acknowledgments}
The John Templeton Foundation is thanked for its generous support of this research. The research was also supported by the Perimeter Institute for Theoretical Physics. The Perimeter Institute was supported by the Government of Canada through Industry Canada and by the Province of Ontario through the Ministry of Economic Development and Innovation. We thank Niayesh Afshordi, Bahram Mashhoon, Martin Green and Viktor Toth for helpful discussions and comments. Finally, we would like to thank David Weinberg ``as referee of this work'' for useful comments and suggestions for improving this work. 
\bibliographystyle{mn2e}

\end{document}